\pdfoutput=1
\documentclass[twocolumn,a4paper]{article}
\usepackage[utf8]{inputenc}
\usepackage{lmodern}
\usepackage{fullpage}
\usepackage{abstract}
\usepackage[UKenglish]{babel}
\usepackage[hidelinks,pdfusetitle]{hyperref}
\usepackage{graphicx}
\usepackage[font=small,font=sf,labelfont=bf]{caption}
\usepackage{authblk}

\begin{document}

\title{On-chip magnetic cooling of a nanoelectronic device}

\author[1]{D. I. Bradley}
\author[1]{A. M. Guénault}
\author[2]{D. Gunnarsson\protect\footnotemark[3]}
\author[1]{R. P. Haley}
\author[1]{S. Holt}
\author[1]{A. T. Jones}
\author[1]{Yu.\@ A. Pashkin}
\author[3]{J. Penttilä}
\author[1]{J. R. Prance\footnote{Corresponding author. Email: \href{mailto:j.prance@lancaster.ac.uk}{j.prance@lancaster.ac.uk}}}
\author[2]{M. Prunnila\footnote{Corresponding author. Email: \href{mailto:Mika.Prunnila@vtt.fi}{Mika.Prunnila@vtt.fi}}}
\author[3]{L. Roschier\footnote{Present address: BlueFors Cryogenics Oy, Arinatie 10, 00370 Helsinki, Finland.}}
\affil[1]{Department of Physics, Lancaster University, Bailrigg, Lancaster LA1 4YB, UK.}
\affil[2]{VTT Technical Research Centre of Finland Ltd, P.O. Box 1000, 02044 VTT, Espoo, Finland.}
\affil[3]{Aivon Oy, Valimotie 13A, 00380 Helsinki, Finland.}

\date{}

\twocolumn[
\begin{@twocolumnfalse}

\maketitle
\begin{abstract}
We demonstrate significant cooling of electrons in a nanostructure below $10\,\mathrm{mK}$ by demagnetisation of thin-film copper on a silicon chip. Our approach overcomes the typical bottleneck of weak electron-phonon scattering by coupling the electrons directly to a bath of refrigerated nuclei, rather than cooling via phonons in the host lattice. Consequently, weak electron-phonon scattering becomes an advantage. It allows the electrons to be cooled for an experimentally useful period of time to temperatures colder than the dilution refrigerator platform, the incoming electrical connections, and the host lattice. There are efforts worldwide to reach sub-millikelvin electron temperatures in nanostructures to study coherent electronic phenomena and improve the operation of nanoelectronic devices. On-chip magnetic cooling is a promising approach to meet this challenge. The method can be used to reach low, local electron temperatures in other nanostructures, obviating the need to adapt traditional, large demagnetisation stages. We demonstrate the technique by applying it to a nanoelectronic primary thermometer that measures its internal electron temperature. Using an optimised demagnetisation process, we demonstrate cooling of the on-chip electrons from $9\,\mathrm{mK}$ to below $5\,\mathrm{mK}$ for over 1000 seconds.
\end{abstract}
\bigskip
\end{@twocolumnfalse}
]
\saythanks

\section*{Introduction}

Nuclear demagnetisation refrigeration is a technique that has been used for several decades to access the $\mathrm{\mu K}$ regime in bulk materials~\cite{Pobell2007}. In practice it requires an elaborate demagnetisation stage to be attached to a dilution refrigerator~\cite{Pickett2000}, which is typically cooled by liquid cryogens. The technique has recently been transferred to cryogen-free dilution refrigerators to cool both $\mathrm{PrNi_5}$~\cite{Batey2013} and $\mathrm{Cu}$~\cite{Todoshchenko2014} refrigerants, requiring additional efforts to compensate for the vibrations present in cryogen-free systems. Here we perform nuclear demagnetisation refrigeration directly on the chip of an aluminium nanoelectronic circuit plated with a thick copper film. This approach removes the requirement for a separate demagnetisation stage and does not require additional vibration isolation. In reaching for lower electron temperatures, we are motivated by the potential improvement in the operation of nanoelectronic devices, for example ensuring charge pumps~\cite{Pekola2013} operate with a minimized leakage and back-tunnelling rate~\cite{Nakamura2014a}, improving the signal to noise ratio of single-electron transistor based sensors~\cite{Giazotto2006} and reducing the charge noise in qubits~\cite{Pashkin2009}.

The nanoelectronic device used in this work, a Coulomb Blockade Thermometer (CBT), was chosen because it provides primary thermometry of the refrigerated electrons. A CBT consists of an array of metallic islands connected by tunnel junctions and operates in the weak Coulomb blockade regime where the charging energy of the islands $E_{c} \ll k_{B}T$. The operating temperature can be tuned by selecting the total island capacitance $C_{\Sigma}$~\cite{Pekola1994}. CBTs have been designed to work at temperatures as high as $60\,\mathrm{K}$, to assist with the redefinition of the Kelvin based on fundamental physical constants~\cite{Meschke2016}. Devices designed for ultralow temperature operation have been tested down to record low temperatures reaching $3.7\,\mathrm{mK}$~\cite{Bradley2016}. Below $10\,\mathrm{mK}$ it becomes challenging to cool the electrons in the CBT as they become increasingly decoupled from the cold lattice, and to overcome this limitation we demonstrate an on-chip cooling method based on nuclear demagnetization of the CBT islands.

On-chip electron refrigeration has a 20 year history starting from the early works on normal metal-insulator-superconductor~\cite{Leivo1997,Nahum1994} and semiconductor-superconductor junction coolers~\cite{Savin2001}. Continuous improvements in the thermal isolation and advances in the junction technology~\cite{ONeil2012,Nguyen2014,Gunnarsson2015} are taking these technologies towards practical on-chip cooling platforms~\cite{Lowell2013}. In addition to the junction coolers, semiconductor quantum dots~\cite{Prance2009} and demagnetisation of Mn-doped aluminium~\cite{Ciccarelli2016} have been investigated for on-chip cooling. In the technique we describe here, the weak electron-phonon coupling in the CBT islands is used as an intrinsic heat switch to pre-cool the electrons. Once the electron-phonon coupling becomes too weak for effective external cooling, we then use demagnetisation to cool the chip from the inside-out while using the CBT’s reliable thermometry to monitor the process. We use a proven low-temperature refrigerant (copper) to extend the range of on-chip refrigeration to significantly lower temperatures than the operating range of most commercial dilution refrigerators.

\section*{Results}
\label{sec:results}
\subsection*{Structure and heatsinking of devices}
\label{sec:results:struct}
\begin{figure}[h!]
  \centering
    \includegraphics{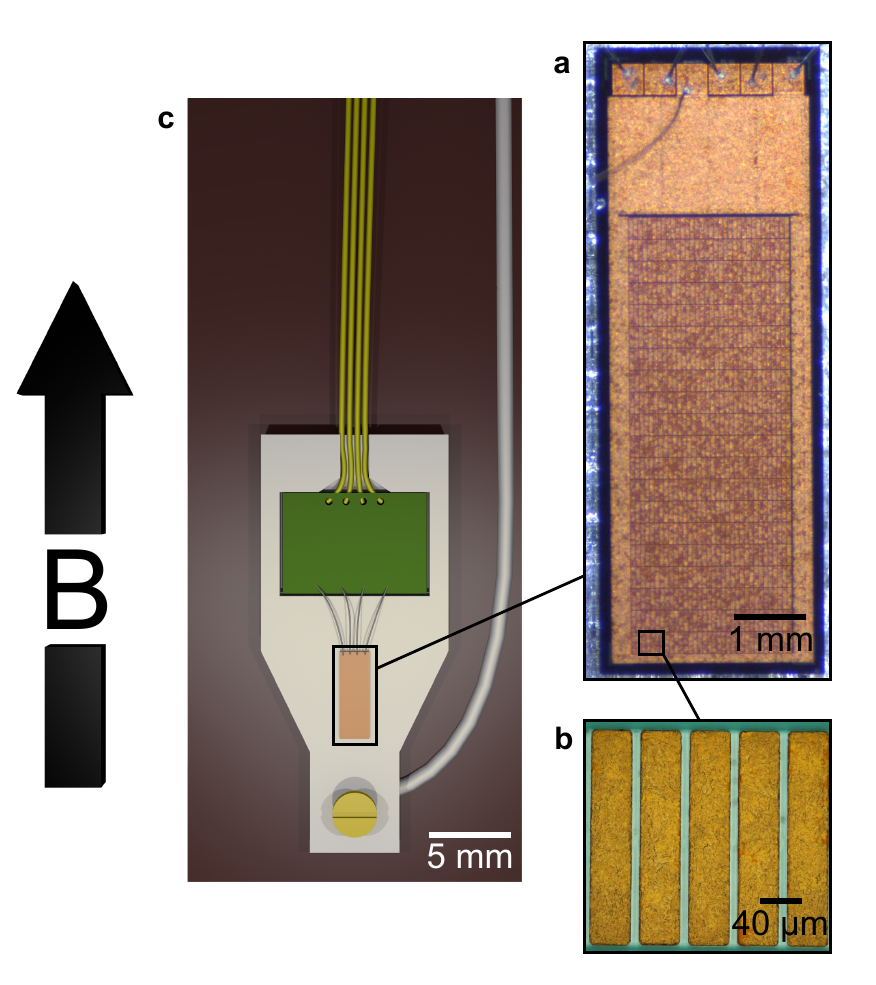}
  \caption{Details of the CBT device structure and package. The large arrow shows the direction of applied magnetic field $B$. Panel (\textbf{a}) is a photograph of the CBT chip, showing the array of $20 \times 32$ copper islands in the bottom $3/4$ of the image and the on-chip filters at the top. Panel (\textbf{b}) is a micrograph showing five of the islands. The package is located in the bore of a superconducting solenoid and the magnetic field is aligned parallel to the long axis of the islands. Panel (\textbf{c}) shows the design of the sample package used to allow effective precooling of the CBT device. It consists of a closed silver box (lid not shown here) in which the CBT is bonded in place using a silver conductive coating (Electrodag\textregistered). The package also contains a small PCB (green) with RLC low-pass filters to which the CBT bond wires are attached. For thermalisation, the package is mounted on a thin copper strip and attached to a silver wire, both of which are attached to the mixing chamber plate of the dilution refrigerator. Four copper wires pass from the package to additional filters at the mixing chamber plate.}
  \label{fig:struct}
\end{figure}

The CBT device, as shown in Fig.\@~\ref{fig:struct}(a), consists of 20 electrically parallel rows of 33 tunnel junctions in series. This arrangement creates a $20 \times 32$ array of metallic islands which increases the accuracy of temperature measurements~\cite{Pekola2000} while also providing a conveniently measurable resistance. The $\mathrm{Al / Al_2O_3}$ junctions, each with resistance $R_{T}$, were fabricated using an ex situ tunnel junction process~\cite{Prunnila2010} (see materials and methods for details). The islands between each pair of junctions are electroplated with copper and have dimensions $39 \times 206 \times 6.14\,\mathrm{\mu m ^3}$ ($W \times L \times T$), as shown in Fig.\@~\ref{fig:struct}(b). To suppress superconductivity of aluminium in the device, a magnetic field of at least $0.1\,\mathrm{T}$ is applied to the CBT during all the measurements presented here.

The $2.3 \times 6.5\,\mathrm{mm^2}$ rectangular chip of the device is contained in a silver package, as shown in Fig.\@~\ref{fig:struct}(c). The package is designed to maximize the thermal conductance between the CBT and the mixing chamber plate of the dry dilution refrigerator (Bluefors Cryogenics LD250). The design also aims to reduce eddy current heating during changes of magnetic field by minimizing the cross-sectional area of metal perpendicular to the field. The package contains electronic filters to provide a quiet electromagnetic environment for the CBT (see materials and methods for more details).

\subsection*{Electron thermometry in constant magnetic fields}
\label{sec:results:therm}
In all the experiments described below, the CBT is measured in a four terminal configuration driven by a custom voltage-controlled current source. Two voltage sense wires from the CBT connect to a differential preamplifier and then to a lock-in amplifier and voltmeter. The current source is used to apply an AC (around $80\,\mathrm{Hz}$) excitation of $40\,\mathrm{pA}$ and a controllable DC bias current. The lock-in and voltmeter measure the differential conductance of the CBT and the DC bias voltage, respectively.

The electron temperature $T_{e}$ in the CBT can be determined in two different ways. In the primary mode, the differential conductance of the CBT is measured for a range of DC bias values and a dip in conductance is observed around zero bias. The dip in conductance becomes deeper and sharper at lower temperatures. In the well-thermalised regime, the full width at half maximum $V_{1/2}$ of the dip is related to temperature by $V_{1/2} \approx 5.439 N k_{B} T / e$, where $N$ is the number of tunnel junctions in series (here $N = 33$)~\cite{Pekola1994}. The CBT can also be operated in secondary mode~\cite{Bradley2016}, whereby only the conductance at zero bias is measured. This is much faster than measuring the full conductance curve and the electron temperature can be determined from a pre-calculated lookup table of zero bias conductances. The lookup table is calculated from the sensor parameters $C_{\Sigma}$ and $R_{T}$ which in turn can be found from a self-calibration procedure described below. The self-calibration and the production of the lookup table were performed using code based on the free, open source python library pyCBT.

\begin{figure*}[ht!]
  \noindent\makebox[\textwidth]{%
  \centering
    \includegraphics{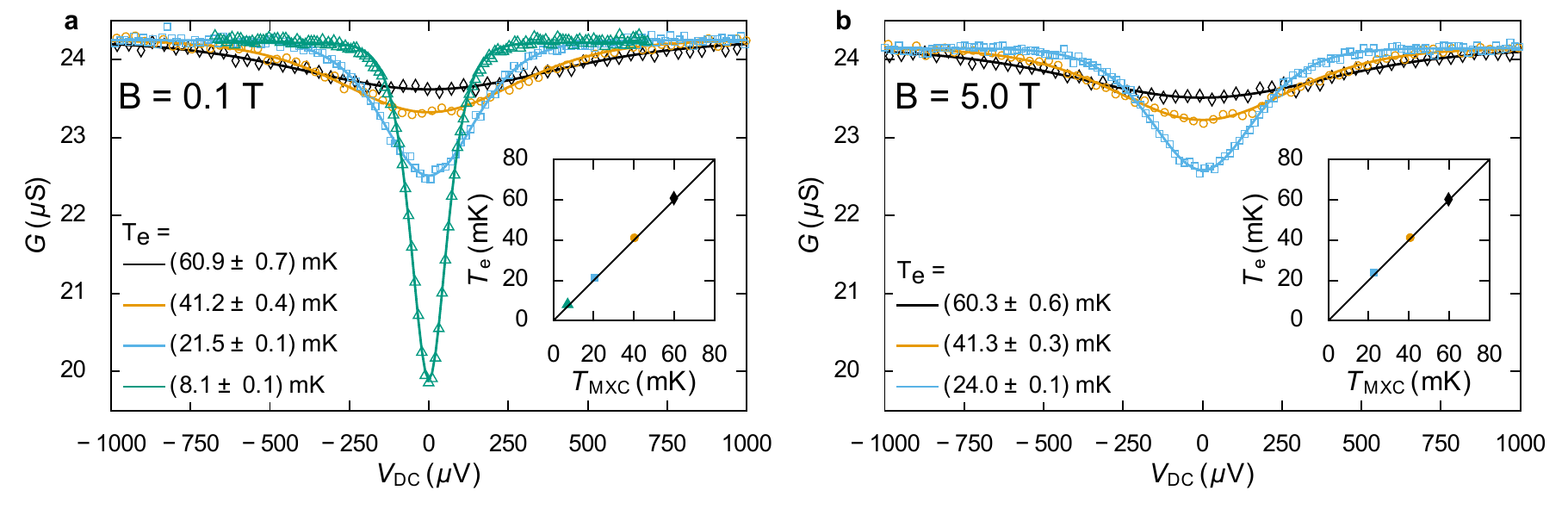}}
  \caption{Calibration of the CBT at different magnetic fields. Panel (\textbf{a}) shows the CBT conductance $G$ against measured DC bias $V_{DC}$ for several different temperatures at $0.1\,\mathrm{T}$. Panel (\textbf{b}) shows similar measurements made at $5.0\,\mathrm{T}$. In both panels, the symbols show measured values and the lines show the best fit to the modelled conductance. The warmest three measurements (black diamonds, orange circles and blue squares) are fitted simultaneously to find the island capacitance $C_{\Sigma}$ and junction resistance $R_{T}$. The parameters found were $C_{\Sigma} = 192.4 \pm 0.9\,\mathrm{fF}$ and $R_{T} = 24.99 \pm 0.06\,\mathrm{k\Omega}$ at $0.1\,\mathrm{T}$, and $C_{\Sigma} = 191.9 \pm 0.8\,\mathrm{fF}$ and $R_{T} = 25.10 \pm 0.06\,\mathrm{k\Omega}$ at $5.0\,\mathrm{T}$. The legends show the fitted electron temperatures (uncertainties from the confidence in the fitting parameters). These values are also plotted in the insets against the temperature of the dilution refrigerator’s mixing chamber plate $T_\mathrm{MXC}$ as measured by a $\mathrm{RuO_2}$ thermometer. The solid line in the insets shows $T_{e} = T_\mathrm{MXC}$, and it can be seen that the CBT electron temperature remains well-thermalised with the mixing chamber. A conductance curve was measured at base temperature (green diamonds in panel (\textbf{a})) after the CBT was allowed to cool for 5 hours. From this we find a base electron temperature of $8.1 \pm 0.1\,\mathrm{mK}$ at a fridge temperature $T_\mathrm{MXC} = 7.12 \pm 0.06\,\mathrm{mK}$.}
  \label{fig:cal}
\end{figure*}

The CBT is self-calibrated at magnetic fields of $0.1\,\mathrm{T}$ and $5.0\,\mathrm{T}$ by measuring conductance curves at three different temperatures for DC biases up to $\pm 1\,\mathrm{mV}$. By fitting these curves simultaneously to a full tunnelling model of the CBT conductance, the values of the total capacitance $C_{\Sigma}$ and tunnel resistance $R_{T}$ can be found~\cite{Bradley2016} as well as the values of the three temperatures. This fitting is performed for the three highest temperature curves in Fig.\@~\ref{fig:cal}(a) and \ref{fig:cal}(b), and we find that the parameters $C_{\Sigma}$ and $R_{T}$ are independent of field, within the experimental error, between $5.0\,\mathrm{T}$ and $0.1\,\mathrm{T}$.

The exact temperatures of the conductance curves are not important for the purposes of calibration, it is only required that the measurements are made at sufficiently different temperatures for $C_{\Sigma}$ and $R_{T}$ to be fitted with a low degree of uncertainty. We confirm that the fits are good by comparing the extracted values of electron temperature to the temperature of the mixing chamber $T_\mathrm{MXC}$, which was monitored using a $\mathrm{RuO_2}$ thermometer, supplied and calibrated by Bluefors Cryogenics (see Fig.\@~\ref{fig:cal}).

The fitting of the three curves at $0.1\,\mathrm{T}$, shown in Fig.\@~\ref{fig:cal}(a), yields sensor parameters $C_{\Sigma} = 192.4 \pm 0.9\,\mathrm{fF}$ and $R_{T} = 24.99 \pm 0.06\,\mathrm{k\Omega}$. At $5.0\,\mathrm{T}$, in Fig.\@~\ref{fig:cal}(b), the parameters are $C_{\Sigma} = 191.9 \pm 0.8\,\mathrm{fF}$ and $R_{T} = 25.10 \pm 0.06\,\mathrm{k\Omega}$. The uncertainties in these parameters arise from the confidence in the fitting, the experimental random error and the drift in gain of the preamplifier and lock-in input amplifier. The averages of these sensor parameters are then used to calculate the lookup table required for secondary mode operation. The uncertainty in the measured parameters corresponds to an uncertainty in electron temperature of less than $1\,\%$ at $5\,\mathrm{mK}$.

Following the $0.1\,\mathrm{T}$ calibration the CBT was allowed to thermalise at base temperature ($T_\mathrm{MXC} = 7.1\,\mathrm{mK}$) for around 5 hours to account for the long thermalisation times observed previously for a CBT in vacuum~\cite{Bradley2016}. A further conductance curve was then measured (see Fig.\@~\ref{fig:cal}(a)) and fitted to find the base electron temperature of $8.1\,\mathrm{mK}$.

\subsection*{Demagnetisation cooling}
\label{sec:results:demag}
To magnetise and demagnetise the on-chip islands, the CBT package is located inside the bore of a superconducting solenoid with a maximum field of $5\,\mathrm{T}$. For all demagnetisation experiments, the magnetic field is initially increased to $5\,\mathrm{T}$, during which the heat of magnetisation and eddy current heating raise the CBT electron temperature. The field is then held constant until the fridge and electrons in the CBT are thermalised, taking approximately 14 hours. Demagnetisation is then performed by sweeping the magnetic field down to the chosen value (not less than $0.1\,\mathrm{T}$) while measuring the CBT in secondary mode approximately every six seconds. 

\begin{figure*}[ht!]
  \centering
    \includegraphics{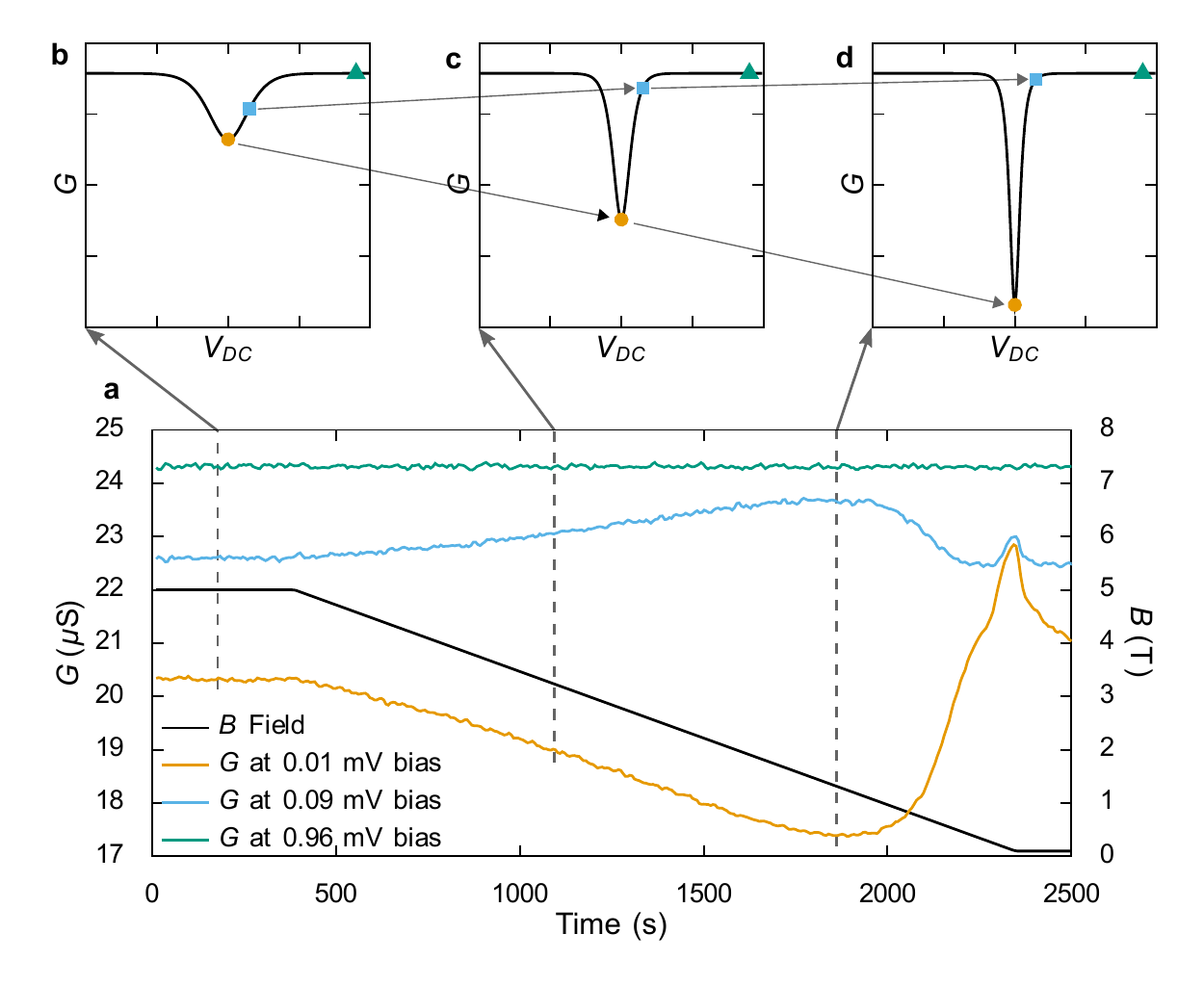}
  \caption{Variation of CBT conductance with DC bias during demagnetisation. In panel (\textbf{a}), the solid black line shows the magnetic field sweeping at $2.5\,\mathrm{mT/s}$ from $5.0$ to $0.1\,\mathrm{T}$. The orange, blue and green lines show the peak, half-maximum and asymptotic CBT conductances, respectively. Panels (\textbf{b}), (\textbf{c}) and (\textbf{d}) are schematic CBT conductance characteristics (plotted against common scales) for reducing electron temperature from (\textbf{b}) to (\textbf{d}). The coloured symbols show the three conductances at the indicated times.}
  \label{fig:demagbias}
\end{figure*}

Fig.\@~\ref{fig:demagbias}(a) shows the CBT conductance during a demagnetisation from $5.0\,\mathrm{T}$ to $0.1\,\mathrm{T}$ at a rate of $2.5\,\mathrm{mT/s}$. The conductance at zero bias is reduced by around $15\,\%$ during the magnetic field sweep, consistent with the electron temperature decreasing and hence the conductance curve peak becoming sharper. To verify this interpretation, the change in CBT conductance during demagnetisation was measured at different DC bias values, as shown in Fig.\@~\ref{fig:demagbias}(a). The applied DC bias was set to $0.01\,\mathrm{mV}$, $0.09\,\mathrm{mV}$ and $0.96\,\mathrm{mV}$ which correspond to the centre of the conductance dip, the half maximum point of the conductance dip (when measured prior to demagnetisation) and the asymptotic conductance, respectively. The conductance curves at the different biases show completely different trends, but all three are consistent with the conductance dip becoming deeper and sharper, corresponding to a reduction in electron temperature. We also note that the asymptotic conductance throughout the demagnetisation takes the value $24.31 \pm 0.03\,\mathrm{\mu S}$, the standard deviation of which ($0.12\,\%$) is smaller than the previously assessed experimental error in tunnel resistance ($0.24\,\%$), meaning the asymptotic conductance is constant during a demagnetisation within the experimental error. This implies the observed drop in zero-bias conductance is not an artefact from, for example, induced voltages and is indeed consistent with cooling of the CBT island electrons.

\begin{figure*}[ht!]
  \noindent\makebox[\textwidth]{%
  \centering
    \includegraphics{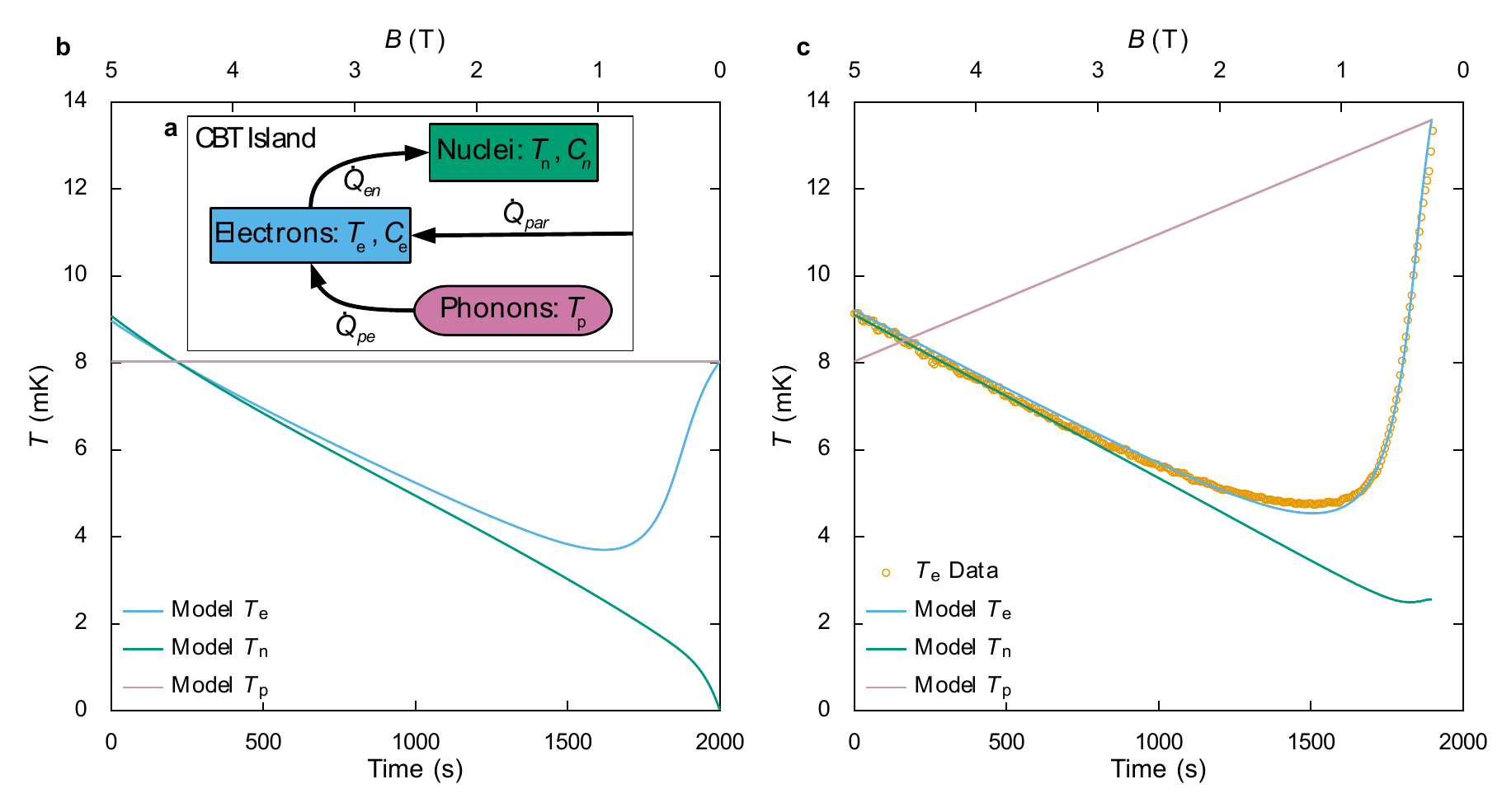}}
  \caption{Numerical model of demagnetisation refrigeration. Panel (\textbf{a}) shows a schematic of the thermal model. Here the CBT island electrons (measured temperature $T_{e}$ and heat capacity $C_{e}$) are connected to the island phonons (temperature $T_{p}$) through electron-phonon coupling, resulting in a heat flow $\dot{Q}_{pe}$. They are also connected to the CBT island nuclei (temperature $T_{n}$ and heat capacity $C_{n}$) with a heat flow $\dot{Q}_{en}$ and have an additional parasitic heat leak $\dot{Q}_{par}$. Panel (\textbf{b}) is a simulation of an ideal demagnetisation with no external heat leak and a constant phonon temperature. Panel (\textbf{c}) is a fitting of this model to the $2.5\,\mathrm{mT/s}$ demagnetisation shown in Fig.\@~\ref{fig:demagbias}. For this fitting, the measured electron temperatures $T_{e}$ (orange circles) were fitted to the calculated electron temperatures by varying $\dot{Q}_{par}$, the rate of increase of $T_{p}$ and the electron-phonon coupling effective volume. The fitting parameters $\dot{Q}_{par}$ and the rate of increase of $T_{p}$ are $6.317\,\mathrm{fW}$ and $2.960\,\mathrm{\mu K/s}$, respectively, and the effective volume used to calculate $\dot{Q}_{pe}$ is $12\,\%$ of the volume used for $\dot{Q}_{en}$. We have previously found $\dot{Q}_{par} > 0.3\,\mathrm{fW}$ for similar devices measured in a lower noise environment \cite{Bradley2016}.}
  \label{fig:demagmodel}
\end{figure*}

To further understand the demagnetisation process we model the heat flow in the CBT using three coupled sub-systems within each copper island: the nuclei, electrons and phonons, see Fig.\@~\ref{fig:demagmodel}(a). We assume that the island phonons are coupled to the substrate, and that this thermal bath has a much larger heat capacity than the electrons and nuclei. The heat flow from the phonon bath to the electrons due to electron-phonon coupling is given by~\cite{Wellstood1994,Giazotto2006}
\begin{equation} \label{eq:qpe}
\dot{Q}_{pe} = \Sigma V \left( T_{p}^5 - T_{e}^5 \right),
\end{equation}
where $\Sigma$ is a material dependent electron-phonon coupling constant, measured previously as $2\,\mathrm{GW/(m^3K^5)}$~\cite{Giazotto2006,Viisanen2016}, and $V$ is the CBT island volume. The heat flow between the island electrons and nuclei is~\cite{Huiskamp1973}
\begin{equation} \label{eq:qen}
\dot{Q}_{en} = \frac{\lambda n \left( B^2 + b^2 \right)}{\mu_0 \kappa T_{n}} \left(T_{e} - T_{n}\right),
\end{equation}
where $\lambda$ is the nuclear Curie constant, $n$ the number of moles of copper on the island, $b = 0.36\,\mathrm{mT}$ the effective dipolar field in copper~\cite{Pobell2007}, $\mu _0$ the permeability of free space and $\kappa = 1.2$ the Korringa constant for copper~\cite{Enss2005}. In addition, we include a constant parasitic heat leak $\dot{Q}_{par}$ into the island electrons that may be due to electronic noise in the measurement circuit or non-equilibrium photons in the environment.

We assume that the thermalisation processes between the sub-systems (i.\@e.\@ electron-nuclear and electron-phonon coupling) are much slower than the internal thermalisation of the nuclear system. We can therefore model the demagnetisation as small, instantaneous, adiabatic steps of magnetic field separated by periods of thermalisation between the sub-systems. During each magnetic field step $\delta B$, the nuclear temperature changes by an amount $\delta T_{n}$. For an ideal adiabatic step $B/T_{n}$ is constant~\cite{Pobell2007}, giving
\begin{equation} \label{eq:deltat}
\delta T_{n} = T_{n} \frac{\delta B}{B}.
\end{equation}
Between the magnetic field steps, the electron temperature evolves according to
\begin{equation} \label{eq:dtedt}
\frac{dT_{e}}{dt} = \frac{\dot{Q}_{pe} + \dot{Q}_{par} - \dot{Q}_{en}}{C_{e}},
\end{equation}
and the nuclear temperature evolves according to
\begin{equation} \label{eq:dtndt}
\frac{dT_{n}}{dt} = \frac{\dot{Q}_{en}}{C_{n}} = \frac{T_{n}}{\kappa}\left(T_{e} - T_{n} \right).
\end{equation}
Here $C_{n}$ and $C_{e}$ are the nuclear and electronic heat capacities, respectively, given by
\begin{equation} \label{eq:cn}
C_{n} = \frac{\lambda n\left( B^2 + b^2 \right)}{\mu _0 T_{n}^2},
\end{equation}
and
\begin{equation} \label{eq:ce}
C_{e} = \frac{\pi ^2}{2} n R \frac{T_{e}}{T_{F}},
\end{equation}
where $T_{F} = 8.12 \times 10^4\,\mathrm{K}$ is the Fermi temperature for copper~\cite{Pobell2007} and $R$ the ideal gas constant.

Solving these equations numerically for the ideal case of $\dot{Q}_{par} = 0$ and constant $T_{p}$ gives the demagnetisation temperature curves shown in Fig.\@~\ref{fig:demagmodel}(b). Even in this simple case, the model reproduces the most significant features seen in the data: an initial drop of $T_{e}$ followed by a sharper increase of $T_{e}$ below $\approx 1\,\mathrm{T}$. In order to achieve a close quantitative agreement with the data, it is necessary to introduce three parameters that modify the model to more closely resemble the physical situation. First, we allow $\dot{Q}_{par}$ to be a fitting parameter representing constant background heating from the environment. Second, we decouple the effective island volumes that are used to calculate $\dot{Q}_{pe}$ and $\dot{Q}_{en}$. This allows the model to account for differences in the heat capacities of the electroplated copper material on the CBT and the high purity copper used to determine $\Sigma$ in~\cite{Giazotto2006,Viisanen2016} and $\lambda$ and $\kappa$ in~\cite{Pobell2007}. Finally, we allow the phonon temperature $T_{p}$ to increase linearly during the demagnetisation. We expect $T_p$ to increase due to eddy current heating of the sample package, however the actual change in phonon temperature cannot be measured directly by the CBT. We fit a linear increase in $T_p$ as the simplest approximation for phonon heating. This increase also allows the model to account for differences between the exact functional form of the electron-phonon coupling and that given by equation~\ref{eq:qpe}. Deviations from a $T^5$ dependence have been reported in similar devices~\cite{Bradley2016,Casparis2012}. Both eddy current heating and any deviation from equation~\ref{eq:qpe} can be fitted using a linear increase in $T_p$. In our experiment, both effects are present simultaneously and it is not possible to separate them. As such it is not possible to determine to what extent $T_p$ reflects the true phonon temperature during a demagnetisation. In the fitting, the initial value of $T_p$ is set to the temperature of the mixing chamber plate, as measured by the $\mathrm{RuO_2}$ thermometer. By fitting the three parameters discussed above, we find close agreement between the calculated $T_e$ and the data for five different demagnetisations at different sweep rates. Figure~\ref{fig:demagmodel}(c) shows the fit to a demangetisation made at $2.5\,\mathrm{mT/s}$.

Using the thermal model, we optimize the demagnetisation process to increase the low temperature hold time. From equation~\ref{eq:qen} it can be seen that the heat transfer from the electrons to the nuclei goes as $B^2$, meaning that the cooling power is very significantly reduced towards the end of the demagnetisation even though $T_{n}$ is significantly lower than $T_{e}$, as shown in Fig.\@~\ref{fig:demagmodel}(b). This means that the electron-phonon coupling, which grows stronger (equation~\ref{eq:qpe}) due to the falling $T_{e}$ and rising $T_{p}$, becomes dominant and rapidly heats the electrons. Reducing the magnetic field ramp rate at the lower field values minimises eddy current heating, hence the rise in $T_{p}$, and keeps the field at higher values for longer, allowing more heat to flow between the electrons and nuclei. Furthermore, stopping the demagnetisation at higher fields stops any eddy current heating before the heat capacity of the nuclei (equation~\ref{eq:cn}) has become negligibly small.

\begin{figure*}[ht!]
  \centering
    \includegraphics{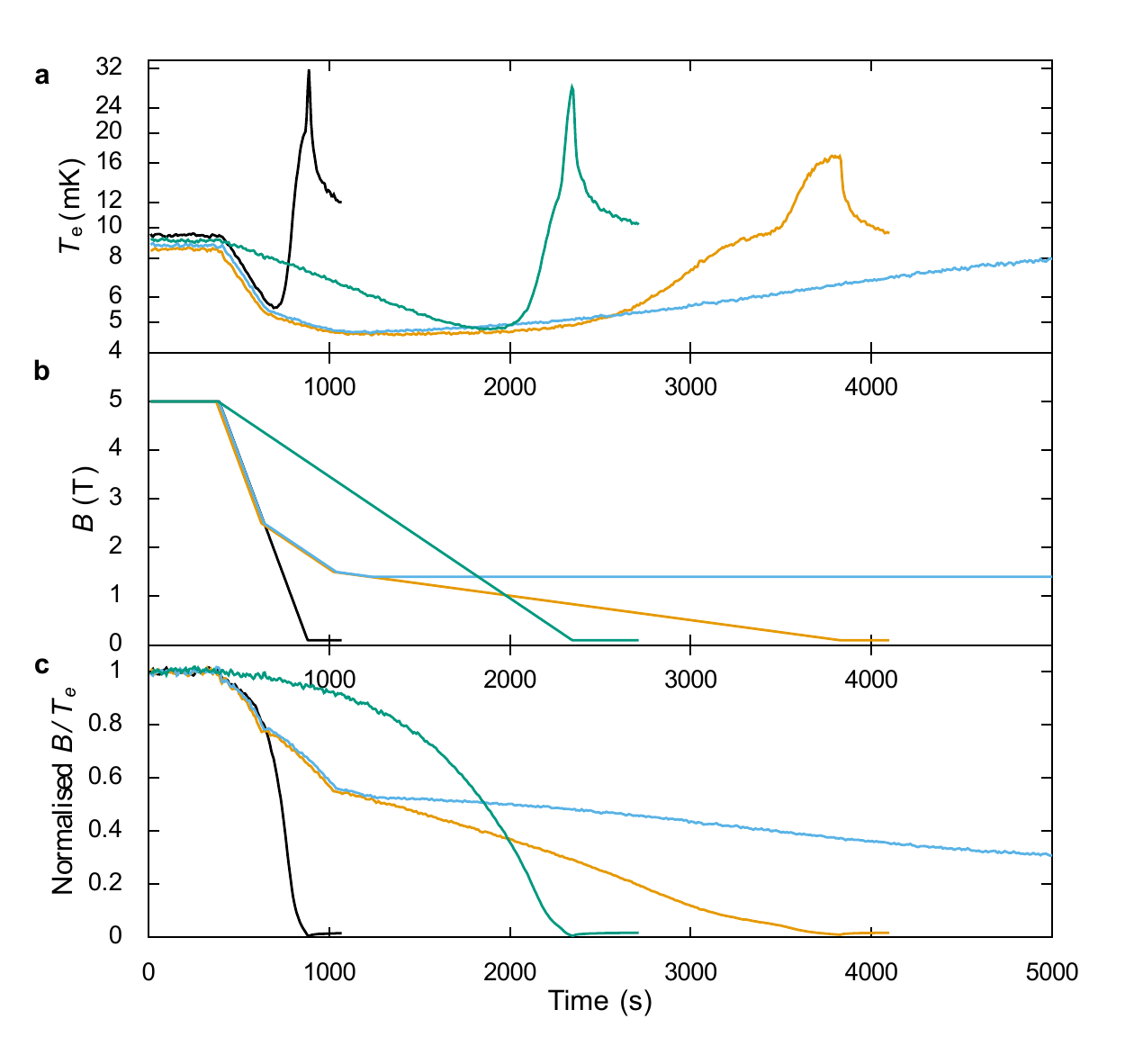}
  \caption{Dependence of demagnetisation cooling on field ramp profile. Panel (\textbf{a}) shows the variation in CBT electron temperature with time for the field profiles shown in panel (\textbf{b}) for the corresponding colours. All profiles proceeded from $5.0\,\mathrm{T}$. The black and green profiles ramp to $0.1\,\mathrm{T}$ at constant rates of $10\,\mathrm{mT/s}$ and $2.5\,\mathrm{mT/s}$, respectively. The other profiles ramp at $10\,\mathrm{mT/s}$ to $2.5\,\mathrm{T}$, then at $2.5\,\mathrm{mT/s}$ to $1.5\,\mathrm{T}$ and finally at $0.5\,\mathrm{mT/s}$ to $0.1\,\mathrm{T}$ and $1.4\,\mathrm{T}$ for the orange and blue curves, respectively. Panel (\textbf{c}) shows the variation in $B/T_{e}$ during these demagnetisations and hence the amount of deviation from the ideal case of constant entropy (constant $B/T_{e}$).}
  \label{fig:demagprofile}
\end{figure*}

The electron temperatures measured during four different demagnetisation sweeps are shown in Fig.\@~\ref{fig:demagprofile}(a), with the magnetic field profiles shown in Fig.\@~\ref{fig:demagprofile}(b). During an ideal adiabatic demagnetisation there is no entropy change and $B/T_{e}$ is constant~\cite{Pobell2007}. Fig.\@~\ref{fig:demagprofile}(c) shows the normalized values of $B/T_{e}$ for each measurement, indicating how close each is to the ideal case. The $10\,\mathrm{mT/s}$ demagnetisation (black curve) shows a large change in $B/T_{e}$ and the electron temperature also peaks far above the initial temperature at the end of the ramp. Reducing the sweep rate to $2.5\,\mathrm{mT/s}$ reduces this peak and also reduces the lowest temperature achieved to $4.7\,\mathrm{mK}$. However, the time taken to reach the minimum temperature is long and the entropy loss at low temperatures is still large, which causes the electrons to quickly heat back up. Moving to a multi-rate demagnetisation increases the hold time below $5\,\mathrm{mK}$ to $1200\,\mathrm{s}$ (from around $400\,\mathrm{s}$) and reduces the minimum temperature to $4.5\,\mathrm{mK}$. This was further improved by stopping the sweep at $1.4\,\mathrm{T}$ which fully eliminates the peak in electron temperature seen at low fields while still reaching the same ultimate base temperature.

\section*{Discussion}
\label{sec:diss}
We have shown that a CBT with large, copper electroplated islands is a reliable thermometer of electron temperature over a large range of magnetic fields. It also exhibits long thermalisation times below $10\,\mathrm{mK}$, as has been observed previously in devices with gold-plated islands~\cite{Bradley2016}. It is expected that the CBT should be unaffected by magnetic fields as its operation is solely based on electrostatic properties~\cite{Pobell2007}. Our results are in agreement with this expectation and with a previous investigation into the magnetic field dependence of CBTs~\cite{Hirvi1996}.

When the Cu islands of the CBT are demagnetised we see that the minimum CBT conductance falls while the asymptotic conductance remains constant and the intermediate conductance increases, as shown in Fig.\@~\ref{fig:demagbias}(a). This is due to the sharpening and deepening of the dip in the conductance curve and is direct evidence of cooling. We model this cooling process by considering the flow of heat between three subsystems in each island, and the model fits closely to the experimentally measured electron temperatures. One notable deviation is that the lowest temperatures found in the data are consistently slightly higher than those found by the numerical simulation, for example as seen in Fig.\@~\ref{fig:demagmodel}(c). This is consistent with electrical noise in the measurement system causing a small amount of conductance curve broadening (see materials and methods) and hence overestimated electron temperatures in the data. Significant efforts were made to minimise noise sources, but we cannot be certain that noise broadening is not limiting the lowest temperatures reported.

We have modified the magnetic field sweep profiles to prevent the electron temperature peaking above the initial temperature, to minimize the lowest temperature reached and to maximize the time the CBT remains at the lower temperatures. Using these profiles, the data show a reduction in minimum temperature that is relatively small (less than $1\,\mathrm{mK}$), but a dramatic increase in the time that the CBT remains below $5\,\mathrm{mK}$ from $\approx 400\,\mathrm{s}$ to $\approx 1200\,\mathrm{s}$. This shows that on-chip magnetic refrigeration can be a practical cooling method to allow measurements at temperatures lower than those achievable with commercial dry dilution refrigerators (typically guaranteed around $10\,\mathrm{mK}$).

The thermal model allows us to simulate the demagnetisation process for arbitrary values of magnetic field and starting temperature. Using the best demagnetisation profile from Fig.\@~\ref{fig:demagprofile}, an initial electron temperature of $6\,\mathrm{mK}$ and an initial magnetic field of $10\,\mathrm{T}$, we expect a minimum electron temperature of $1.1\,\mathrm{mK}$. Similarly, with an initial field of $12\,\mathrm{T}$ we expect a minimum temperature of $0.9\,\mathrm{mK}$. These parameters are within reach of commercial dry dilution refrigerators. Finally, we note a previous observation that using low quality copper flakes as a refrigerant leads to surprisingly long thermalisation times after demagnetisation~\cite{Bradley1984}. We believe that this shows potential for further improvements in the low temperature hold time by optimising the material properties of the on-chip copper refrigerant.

In conclusion, we have observed that the use of large copper islands on a Coulomb blockade thermometer allows magnetic cooling of the device and hence permits the electron temperature to be significantly reduced below the temperature of the phonon bath, the temperature of this dilution refrigerator, and the base temperature of most other commercial refrigerators. By optimizing the magnetic field ramp rates, the electrons were cooled below $5\,\mathrm{mK}$ for $1200\,\mathrm{s}$. We anticipate that sub-$1\,\mathrm{mK}$ electron temperatures can be reached using this technique by employing a larger magnetic field and a lower starting temperature.

\section*{Methods}
\label{sec:meth}
\subsection*{CBT fabrication}
\label{sec:meth:cbtfab}
The CBT substrate is an undoped silicon wafer with a $300\,\mathrm{nm}$ thermal oxide layer. On to this, aluminium films ($250\,\mathrm{nm}$ thick) were deposited to form the CBT circuit, with $\mathrm{SiO_2}$ deposited by plasma enhanced chemical vapour deposition to separate different aluminium layers. Tunnel junctions were formed, in an ex situ tunnel junction process~\cite{Prunnila2010}, as an aluminium oxide layer between two aluminium layers contained within a cylindrical via (diameter nominally $0.8\,\mathrm{\mu m}$) through the $\mathrm{SiO_2}$, giving a resistivity of $\sim 10\,\mathrm{k\Omega \mu m^2}$.

The ex situ method, and other typical tunnel junction deposition processes, are generally usable for film thicknesses up to around $1\,\mathrm{\mu m}$. Above this the films can suffer from stress build-up which leads to poor adhesion to the substrate. On cooling to low temperatures this can cause poor thermalisation and, due to the significant thermal contraction during cool down, mechanical failure. To maximize the nuclear heat capacity available in the islands for demagnetisation cooling, it is necessary to have the thickest islands possible. To achieve this, the islands were mask electroplated with copper following the ex situ process. The mask electroplating can be used for low embedded stress films up to $\sim 10\,\mathrm{\mu m}$ thick, and here was used to create an approximately $6\,\mathrm{\mu m}$ thick layer.

\subsection*{CBT packaging}
\label{sec:meth:cbtpak}
To maximize the thermal conductance between the device and the mixing chamber plate, a path of high conductivity silver is used to connect the CBT chip and the plate. This is achieved by bonding the CBT to a silver 3D-printed package using a silver conductive coating (Electrodag\textregistered). A high purity, annealed silver wire of approximately 1 mm diameter connects the outside of the silver package to the gold plated copper mixing chamber plate surface. The ends of the silver wire are crushed between a silver washer and the contact surfaces using a brass bolt, nut and copper washers (used to prevent loosening due to the relatively large thermal contraction of silver on cooling).

The package and silver wire were supported by a plastic rod below the mixing chamber plate of the dilution refrigerator. The choice of a plastic rod was made to minimize the total cross-sectional area of metal perpendicular to magnetic flux, thus minimizing the amount of eddy current heating when sweeping the magnetic field. Similarly, the CBT package is shaped to minimize the cross section perpendicular to the applied field, and the long axis of the copper CBT islands is oriented parallel to the field.

\subsection*{Electronic measurement}
\label{sec:meth:elecmeas}
The measurement circuit includes three stages of electronic filtering to reduce noise at the sample. The effect of electrical noise is to smooth the shape of the measured conductance curve, which would lead to artificially high reported electron temperatures. The filtering involves a distributed $RC$ filter on the CBT itself ($R \approx 500\,\mathrm{\Omega}$, $C \approx 10\,\mathrm{pF}$), a discrete component $RLC$ filter ($R = 4.99\,\mathrm{k\Omega}$, $L = 24\,\mathrm{nH}$, $C = 180\,\mathrm{pF}$) on a neighbouring PCB and a discrete component $RCR$ ($R = 200\,\mathrm{\Omega}$, $C = 1\,\mathrm{nF}$) pi-filter mounted on the mixing chamber plate. This $RCR$ filter is an Aivon Oy Therma that has been modified by potting with copper loaded ($40\,\%$ by mass) Stycast\textregistered 2850GT to minimize heating of the resistors, and to act as a powder filter to attenuate high frequency noise. The inductors making up the $RLC$ filter are oriented with their coil axis perpendicular to the applied magnetic field to prevent induced voltages.

The lock-in technique was used to measure the CBT’s differential conductance. The magnitude of the AC excitation used for this ($40\,\mathrm{pA}$) was chosen to prevent conductance curve broadening while keeping the integration times required for a clean measurement acceptable. Measurements were also made with an excitation of $20\,\mathrm{pA}$; however, this did not show a lower base electron temperature.

\subsection*{Data availability}
\label{sec:meth:data}
All data used in this paper are available at \url{http://dx.doi.org/10.17635/lancaster/researchdata/105} including descriptions of the data sets.

\subsection*{Code availability}
\label{sec:meth:code}
Data analysis was undertaken using custom code derived from the python-based software library ``pyCBT'', which is freely available from Aivon Oy at \url{https://github.com/AivonOy/pyCBT}.

\small \bibliography{}

\section*{Acknowledgements}
\label{sec:ack}
We acknowledge Hannele Heikkinen and Leif Grönberg for assistance in the sample preparation, and thank Jukka Pekola for helpful discussions. This research is supported by the U.K. EPSRC (EP/K01675X/1, EP/N019199/1 and EP/L000016/1), Academy of Finland (projects 252598 and 287768), Tekes project FinCryo (grant number 220/31/2010) and the European FP7 Programme MICROKELVIN (project number 228464). Yu.\@A.P.  acknowledges support by the Royal Society (WM110105). J.R.P. acknowledges support of the People Programme (Marie Curie Actions) of the European FP7 Programme (REA grant agreement 618450).

\end{document}